\documentstyle[prd,aps,epsfig,psfig]{revtex}
\begin{document} 

\tighten
\draft
%\preprint{DAMTP-1999-80}
\twocolumn[\hsize\textwidth\columnwidth\hsize\csname 
@twocolumnfalse\endcsname

\title{Information Content for Quantum States} 

\author{
Dorje C. Brody$^{*}$ 
%\footnote[1]{Electronic address: d.brody@damtp.cam.ac.uk}
and Lane P. Hughston$^{\dagger}$ 
%\footnote[2]{Electronic address: lane_hughston@yahoo.com}
} 

\address{* DAMTP, Silver St., Cambridge CB3 0DS, UK} 
\address{$\dagger$ Department of Mathematics, King's College 
London, The Strand, London WC2R 2LS, UK} 

\date{\today} 

\maketitle 

\begin{abstract}
A method of representing probabilistic aspects of quantum 
systems is introduced by means of a density function on the 
space of pure quantum states. In particular, a maximum 
entropy argument allows us to obtain a natural density function 
that only reflects the information provided by the density matrix. 
This result is applied to derive the Shannon entropy of a quantum 
state. The information theoretic quantum entropy thereby obtained 
is shown to have the desired concavity property, and to differ 
from the the conventional von Neumann entropy. This is 
illustrated explicitly for a two-state system. 
\end{abstract} 

\pacs{PACS Numbers : 03.65.Bz, 05.30.Ch, 02.40.Ft} 

\vskip2pc] 

In standard quantum mechanics, the information about physical 
observables is contained in the {\it state} of the system, which 
is represented by a density matrix $\rho^{\alpha}_{\beta}$. This 
is because the expectation of an observable $F^{\alpha}_{\beta}$ 
in the state $\rho^{\alpha}_{\beta}$ is given by the trace formula 
%%%%%%%%%%%%%%%%%%%%%% 
\begin{equation} 
\langle F\rangle\ =\ \rho^{\alpha}_{\beta}F^{\beta}_{\alpha}, 
\label{eq:exp} 
\end{equation} 
%%%%%%%%%%%%%%%%%%%%%% 
and it is through such expectations that the statistical properties 
of measurement outcomes are determined. Indeed, for a state we 
require $\rho^{\alpha}_{\beta}$ to be nonnegative and to have trace 
unity. These properties suggest that the density matrix can be 
viewed as a probability distribution. For example, if 
$\rho^{\alpha}_{\beta}$ is nondegenerate, with distinct eigenvalues, 
then it admits a unique decomposition 
%%%%%%%%%%%%%%%%%%%%%%%%%%%% 
\begin{equation} 
\rho^{\alpha}_{\beta}\ =\ \sum_{i}w_{i}
\Pi^{\alpha}_{\beta}(x_{i}). 
\label{eq:state} 
\end{equation} 
%%%%%%%%%%%%%%%%%%%%%%%%%%%% 
Here, $\Pi^{\alpha}_{\beta}(x_{i})$ denotes the normalised projection 
operators onto the eigenstates $x_{i}$ of $\rho^{\alpha}_{\beta}$, 
and the corresponding {\it probability weights} $w_{i}$ satisfy 
$w_{i}>0$ and $\sum_{i}w_{i}=1$. Some care has to be taken with this 
interpretation of $\rho^{\alpha}_{\beta}$, because in the present 
context the underpinnings of classical probability are missing, and 
the associated terminology can only be used, therefore, by analogy. 
Nevertheless, von Neumann \cite{jvn}, in pursuit of this 
analogy, was led by a series of ingenious arguments involving 
the thermodynamics of a hypothetical gas of independent systems 
represented by a weighted family of orthogonal pure states, 
to argue that the quantity 
%%%%%%%%%%%%%%%%%%%%%%%% 
\begin{equation} 
S_{\rm vN}\ =\ -\rho^{\alpha}_{\beta} \ln \rho_{\alpha}^{\beta} 
\label{eq:vN} 
\end{equation} 
%%%%%%%%%%%%%%%%%%%%%%%%% 
represents the entropy of the state $\rho^{\alpha}_{\beta}$. In 
the example of the state (\ref{eq:state}), for instance, we have 
$S_{\rm vN}=-\sum_{i}w_{i}\ln w_{i}$, which is the classical 
information entropy associated with the 
probability distribution $w_{i}$. \par 

It is clear, nevertheless, that the von Neumann entropy is 
inadequate for some situations. Suppose, for example, we make a 
measurement of an observable with distinct eigenstates $x_{i}$. 
Then the results of the measurement can be represented 
statistically by the state (\ref{eq:state}), where the weighting 
$w_{i}$ are given by the familiar transition amplitudes taken 
with respect to the initial state. In this case, the entropy of 
the distribution is indeed given by $S_{\rm vN}$, since we know 
that the measurement results in one of the 
eigenstates $x_{i}$ being selected, and that the information 
gained with the knowledge of the outcome precisely 
counterbalances the entropy of the state $\rho^{\alpha}_{\beta}$. 
However, this is a special state of affairs, peculiar to the 
measurement problem, and there is no {\it a priori} justification 
for assuming in general, given $\rho^{\alpha}_{\beta}$, that the 
system is in one or another of the eigenstates of 
$\rho^{\alpha}_{\beta}$. In fact, for a given 
$\rho^{\alpha}_{\beta}$, the implied minimal information 
distribution on the space of pure states is of a more general 
character, as we shall demonstrate in what follows. \par 

In this article we introduce a more realistic formula for the 
entropy of a quantum state. Our expression for the quantum 
entropy is in line with that of Shannon; as a consequence, many 
of the standard results for classical information entropy apply. 
The quantum entropy introduced here differs, in general, from 
the von Neumann entropy. However, like the von Neumann entropy, 
the new entropy can be expressed in terms of the eigenvalues of 
the density matrix, as we shall illustrate explicitly in the case 
of a system characterised by a two dimensional Hilbert space. Our 
methodology has the advantage that it more satisfactorily takes 
into account the significance of information in modern 
quantum theory \cite{caves}. Indeed, whereas von Neumann 
specifically accommodates into his thermodynamic analysis as 
extra information the assumption that the ensemble is composed of 
a weighted system of pure states, each one of which belongs to a 
given complete family of orthogonal pure states, we make no such 
assumption here. Instead, in our approach to the quantum entropy 
problem, we shall be guided by information theoretic principles. \par 

The other ingredient at our disposal, missing in von Neumann's 
theory, is the recognition that the space of pure states in 
quantum theory has the structure of a phase space; that is to 
say, it admits a natural symplectic structure. The quantum phase 
space ${\sl\Gamma}$ is a complex projective space endowed 
with a Hermitian correlation between points and hyperplanes. A 
point $x\in{\sl\Gamma}$ represents a pure state, i.e., an 
equivalence class of wave functions belonging to the 
same ray in Hilbert space. When viewed as a real manifold, 
${\sl\Gamma}$ is known to have a natural Riemannian geometry, 
given by the Fubini-Study metric, which has a compatible 
symplectic structure associated with it \cite{kibble}. A typical 
quantum observable $F^{\alpha}_{\beta}$ is given by a function 
$F(x)$ on ${\sl\Gamma}$ of the form  
%%%%%%%%%%%%%%%%%%%%%%%%%%%4 
\begin{equation} 
F(x)\ =\ \frac{{\bar\psi}_{\alpha}(x)F^{\alpha}_{\beta}
\psi^{\beta}(x)}{{\bar\psi}_{\gamma}(x)\psi^{\gamma}(x)},  
\label{eq:lob} 
\end{equation} 
%%%%%%%%%%%%%%%%%%%%%%%%%%% 
where $\psi^{\alpha}(x)$ denotes any wave function in the 
equivalence class associated with the pure state $x$. With a slight 
departure from the traditional terminology we can refer to the 
function $F(x)$ itself as the observable. Then if $F(x)$ and $G(x)$ 
are observables, their Poisson bracket with respect to the 
symplectic structure is also an observable, given by $i$ times the 
expectation of the commutator of the corresponding operators, taken 
in the pure state $x$. The resulting algebra of quantum observables 
gives ${\sl\Gamma}$ the structure of a Poisson manifold, 
and as a consequence the Schr\"odinger trajectories of pure states 
are given by the integral curves of the symplectic vector field for 
which the generator $H(x)$ is the quantum Hamiltonian. \par  

We shall take the view here that a {\it general} quantum state 
is represented by a density function $\rho(x)$ on ${\sl\Gamma}$, 
satisfying $\rho(x)\geq 0$ and 
%%%%%%%%%%%%%%%%%%%%%%%%%%%5 
\begin{equation} 
\int_{\sl\Gamma}\rho(x)dV\ =\ 1, 
\end{equation} 
%%%%%%%%%%%%%%%%%%%%%%%%%%% 
where $dV$ is the volume element associated with the 
Fubini-Study metric. Thus we can think of $\rho(x)$ as 
an ensemble on the phase space ${\sl\Gamma}$. For 
example, let us consider the measurement of an observable $F(x)$ 
with distinct eigenstate $x_{i}$, when initially the system is in 
a given pure state $x_{0}$. Then for the density function 
corresponding to an ensemble consisting of a large number of 
independent identical copies of the system we can write $\rho(x) 
= \delta(x,x_{0})$ for the initial state, and $\rho(x) = 
\sum_{i}w_{i}\delta(x,x_{i})$ after the measurement has been 
performed. Here $\delta(x,x_{i})$ denotes a delta function on 
${\sl\Gamma}$, concentrated at the point $x_{i}$, and $w_{i}$ is 
the transition amplitude 
between the states $x_{0}$ and $x_{i}$. The expectation of an 
observable $F(x)$ in the general state $\rho(x)$ is then given by 
%%%%%%%%%%%%%%%%%%%%%%%%%%%6 
\begin{equation} 
\langle F\rangle\ =\ \int_{\sl\Gamma}F(x)\rho(x)dV. 
\label{eq:exp2} 
\end{equation} 
%%%%%%%%%%%%%%%%%%%%%%%%%%% 
We can regard (\ref{eq:exp2}) as equating $\langle F\rangle$ with 
the unconditional expectation of the conditional expectation $F(x)$ 
in the pure state $x$. The dynamical evolution of $\rho(x)$ is 
governed by the Liouville equation, where the Poisson bracket 
between $\rho(x)$ and $H(x)$ is determined by the symplectic 
structure on ${\sl\Gamma}$. If $\rho(x)$ is initially given by a 
delta function concentrated on a single pure state, then 
subsequently it remains of that form, and the point of 
concentration follows a Schr\"odinger trajectory. \par 

Now, suppose we introduce the projection operator 
%%%%%%%%%%%%%%%%%%%%%%%%%%%7 
\begin{equation} 
\Pi^{\alpha}_{\beta}(x)\ =\ \frac{{\bar\psi}_{\beta}(x)
\psi^{\alpha}(x)}
{{\bar\psi}_{\gamma}(x)\psi^{\gamma}(x)} 
\end{equation} 
%%%%%%%%%%%%%%%%%%%%%%%%%%% 
corresponding to the pure state represented by a generic point 
$x\in{\sl\Gamma}$. Then, the general quantum state can be expanded 
in terms of its moments \cite{dblh1}. In particular, the lowest 
moment of $\Pi^{\alpha}_{\beta}(x)$ in the state $\rho(x)$ gives 
rise to the density matrix of ordinary quantum mechanics: 
%%%%%%%%%%%%%%%%%%%%%%%%%%%8 
\begin{equation} 
\rho^{\alpha}_{\beta}\ =\ \int_{\sl\Gamma} \rho(x) 
\Pi^{\alpha}_{\beta}(x) dV. 
\label{eq:do} 
\end{equation} 
%%%%%%%%%%%%%%%%%%%%%%%%%%% 
It follows from the formulae above that the expectation (\ref{eq:exp2}) 
agrees with the standard trace formula (\ref{eq:exp}), provided 
$F(x)$ is a linear observable of the form ({\ref{eq:lob}), that 
is, $F(x)=F^{\alpha}_{\beta}\Pi^{\beta}_{\alpha}(x)$. An advantage 
of the general expression (\ref{eq:exp2}) is that it 
can also be applied in the case of a nonlinear observable of the 
Kibble-Weinberg type \cite{wein}. It should be emphasised 
nevertheless that when we consider the statistical properties of 
ordinary linear observables, this 
formulation of quantum mechanics on ${\sl\Gamma}$ is equivalent to 
the conventional Hilbert space approach. \par 
 
Under suitable technical conditions the information in the state 
$\rho(x)$ can be represented by the totality of its moments, and 
a unique expansion of the form 
%%%%%%%%%%%%%%%%%%%%%%%%%%9 
\begin{equation} 
\rho(x) = 1+\mu^{\alpha}_{\beta}\Pi^{\beta}_{\alpha}(x) + 
\mu^{\alpha\alpha'}_{\beta\beta'}\Pi^{\beta}_{\alpha}(x) 
\Pi^{\beta'}_{\alpha'}(x) + \cdots 
\end{equation} 
%%%%%%%%%%%%%%%%%%%%%%%%%%% 
exists, where the $\mu$-coefficients are trace-free and 
totally symmetric. A calculation then shows that the 
$n$-th coefficient is given, up to a combinatorial factor, by 
the trace-free part of the $n$-th moment of 
$\Pi^{\beta}_{\alpha}(x)$. It follows that the density matrix of 
ordinary quantum mechanics in general does not contain all of the 
information about the state of the system. This remains the case 
{\it a fortiori} if we relax the technical conditions and allow 
$\rho(x)$ to belong to a broader class of measures. However, 
if we wish to consider the statistical properties of linear 
observables, then, owing to formula (\ref{eq:exp}), it 
suffices to consider the density matrix exclusively. Because our 
intention here is to investigate the entropy in ordinary quantum 
mechanics, we shall therefore examine the consequences of 
assuming that the information encoded in the density matrix is 
the only information available to us. In this context it is worth 
recalling the work of Mielnik \cite{mie}, who regards the state 
in ordinary quantum theory as an equivalence class of density 
functions each of which gives rise to the same density 
matrix. We note, however, that there is a subtle deficiency in 
his approach, because it treats all distributions that give 
rise to the same density matrix on an equal footing. Clearly, 
some distributions contain more information than others, and 
according to the general principles of information theory we 
must look for the distribution that is least informative, subject 
to the condition that it is consistent with the prescribed 
density matrix. \par  

It should be evident from the foregoing discussion that the 
appropriate expression for the Shannon entropy of a quantum state 
$\rho(x)$ is 
%%%%%%%%%%%%%%%%%%%%%%%%%%%%10 
\begin{equation} 
S_{\rho}\ =\ -\int_{\sl\Gamma}\rho(x)\ln\rho(x)dV. 
\label{eq:ent} 
\end{equation} 
%%%%%%%%%%%%%%%%%%%%%%%%%%%% 
Because $\rho(x)$ is a probability density function defined on 
the smooth manifold ${\sl\Gamma}$, it follows that $S_{\rho}$ 
possesses the standard properties of the Shannon entropy. The 
question we have to address here is thus: given a density matrix 
$\rho^{\alpha}_{\beta}$, how do we express the corresponding 
quantum entropy $S_{\rho}$ in terms of it? Clearly, for a generic 
density matrix, there exist many different density functions 
$\rho(x)$ that give rise to the same $\rho^{\alpha}_{\beta}$. 
Therefore, it is not obvious which $\rho(x)$ we should select. 
This problem can be resolved by recalling our assumption 
that {\it the density matrix is the only 
information available to us}. This implies that the relevant 
density function $\rho(x)$ is the one with minimum information, 
or maximum entropy $S_{\rho}$, subject to the constraint 
(\ref{eq:do}). If we let $\lambda^{\alpha}_{\beta}$ denote the 
Lagrange 
multiplier required for this extremisation problem, then the 
solution is a distribution of the canonical form 
%%%%%%%%%%%%%%%%%%%%%%%%%%%%11 
\begin{equation} 
\rho(x)\ =\ \exp\left( -\lambda^{\alpha}_{\beta} 
\Pi^{\beta}_{\alpha}(x) -\ln Z(\lambda) \right), 
\label{eq:maxent} 
\end{equation} 
%%%%%%%%%%%%%%%%%%%%%%%%%%%% 
where the normalisation is given by the generating function 
%%%%%%%%%%%%%%%%%%%%%%%%%%%%12 
\begin{equation} 
Z(\lambda)\ =\ \int_{\sl\Gamma}\exp\left( 
-\lambda^{\alpha}_{\beta} 
\Pi^{\beta}_{\alpha}(x)\right)dV. 
\label{eq:gf} 
\end{equation} 
%%%%%%%%%%%%%%%%%%%%%%%%%%%% 
The Lagrange multiplier $\lambda^{\alpha}_{\beta}$ is 
determined, up to an arbitrary trace term, by the constraint 
%%%%%%%%%%%%%%%%%%%%%%%%%%%%%13 
\begin{equation} 
-\frac{\partial\ln Z(\lambda)}
{\partial\lambda^{\beta}_{\alpha}}\ =\ \rho^{\alpha}_{\beta}. 
\label{eq:lt} 
\end{equation} 
%%%%%%%%%%%%%%%%%%%%%%%%%%%%% 
The result (\ref{eq:maxent}) is perhaps surprising because in the 
literature of quantum theory the canonical distribution function 
arises typically in the thermal context. \par 

It follows from the expression for the minimum information 
distribution function that the quantum Shannon entropy 
associated with the density matrix $\rho^{\alpha}_{\beta}$ is 
given by a Legendre transformation 
%%%%%%%%%%%%%%%%%%%%%%%%%%%%%14 
\begin{equation} 
S_{\rho}\ =\ \lambda^{\alpha}_{\beta}\rho^{\beta}_{\alpha} 
+ \ln Z(\lambda),  
\label{eq:qsent} 
\end{equation} 
%%%%%%%%%%%%%%%%%%%%%%%%%%%%% 
where $\lambda^{\alpha}_{\beta}$ is determined by the relation 
(\ref{eq:lt}). Alternatively, we can combine (\ref{eq:lt}) and 
(\ref{eq:qsent}) and define $S_{\rho}$ according to the scheme 
%%%%%%%%%%%%%%%%%%%%%%%%%%%%%15 
\begin{equation} 
S_{\rho}\ =\ \sup_{\lambda} \left( \lambda^{\alpha}_{\beta}
\rho^{\beta}_{\alpha} + \ln Z(\lambda)\right) . 
\end{equation} 
%%%%%%%%%%%%%%%%%%%%%%%%%%%%%% 
In fact, one can show that $\ln Z(\lambda)$ is 
convex on the vector space obtained by eliminating the trace of 
$\lambda^{\alpha}_{\beta}$. The argument, as we indicate below, is 
reminiscent of the reasoning used to demonstrate the positivity of 
the heat capacity in statistical mechanics. It follows that 
$\ln Z(\lambda)$ is the convex dual of the entropy, and that 
$S_{\rho}$ is concave over the space of density functions. More 
specifically, we find that 
%%%%%%%%%%%%%%%%%%%%%%%%%%%%%16 
\begin{equation} 
\frac{\partial^{2}\ln Z}{\partial\lambda^{\beta}_{\alpha}
\partial\lambda^{\delta}_{\gamma}} = \int_{\sl\Gamma}\rho(x) 
\left( \Pi^{\alpha}_{\beta} - \rho^{\alpha}_{\beta}\right) 
\left( \Pi^{\gamma}_{\delta} - \rho^{\gamma}_{\delta}\right)dV, 
\label{eq:f-info} 
\end{equation} 
%%%%%%%%%%%%%%%%%%%%%%%%%%%%% 
which shows that the Hessian of $\ln Z(\lambda)$ is given by the 
covariance of the projection operator $\Pi^{\alpha}_{\beta}(x)$, 
which is positive definite for trace-free displacements in the 
value of $\lambda^{\alpha}_{\beta}$. Indeed, the Hessian is 
independent of $\lambda^{\alpha}_{\alpha}$, since under the 
transformation $\lambda^{\alpha}_{\beta}\rightarrow 
\lambda^{\alpha}_{\beta}+\mu \delta^{\alpha}_{\beta}$ we have 
$Z(\lambda)\rightarrow e^{-\mu}Z(\lambda)$. It thus follows that 
(\ref{eq:f-info}) defines a Riemannian metric, known as the 
Fisher-Rao metric, on the parameter space of the 
distribution (\ref{eq:maxent}). Therefore, by convex duality 
\cite{rock}, we conclude that $S_{\rho}$ is concave in the sense 
that if $\rho^{\alpha}_{\beta}(i)$ are density matrices for 
$i=1,2, \cdots,n$ and if $\{w_{i}\}$ is a set of probability 
weights, then 
%%%%%%%%%%%%%%%%%%%%%%%%%%%%%%17 
\begin{equation} 
S_{\rho}\left[ \sum_{i}w_{i}\rho^{\alpha}_{\beta}(i)\right] 
\ \geq\ \sum_{i}w_{i}S_{\rho}\left[\rho^{\alpha}_{\beta}(i)\right], 
\end{equation} 
%%%%%%%%%%%%%%%%%%%%%%%%%%%%%%% 
where $S_{\rho}[\rho^{\alpha}_{\beta}]$ denotes the entropy 
(\ref{eq:qsent}) associated with a given density matrix 
$\rho^{\alpha}_{\beta}$. \par 

This is our main result for the quantum entropy. 
To see that $S_{\rho}$ differs from $S_{\rm vN}$ as a function of 
$\rho^{\alpha}_{\beta}$ we proceed as follows. Suppose, on the 
contrary, that there exists a constant $A$, independent of 
$\rho^{\alpha}_{\beta}$, such that $S_{\rho}=S_{\rm vN}+\ln A$. 
Then solving for $\rho^{\alpha}_{\beta}$ by use of (\ref{eq:vN}) 
and (\ref{eq:qsent}) we obtain $\rho^{\alpha}_{\beta} = A\exp( - 
\lambda^{\alpha}_{\beta})/Z(\lambda)$, which implies that 
$\int_{\sl\Gamma}\Pi^{\alpha}_{\beta}\exp(-\lambda^{\gamma}_{\delta} 
\Pi^{\delta}_{\gamma}(x))dV = A\exp(-\lambda^{\alpha}_{\beta})$ holds 
for all $\lambda^{\alpha}_{\beta}$. Expanding each side to first 
order in $\lambda^{\alpha}_{\beta}$, we reach a contradiction. \par  

We have demonstrated that if the information available at our 
disposal is given solely by the density matrix 
$\rho^{\alpha}_{\beta}$, then the corresponding entropy is given 
by (\ref{eq:qsent}). Conversely, any other form of entropy, 
such as that of von Neumann, implies the knowledge of information 
other than $\rho^{\alpha}_{\beta}$, even if the entropy itself 
can be expressed in terms of $\rho^{\alpha}_{\beta}$. Hence, 
in a strict sense, any other choice of entropy goes beyond 
the category of linear quantum mechanics, as is consistent with 
the fact that the von Neumann entropy gives the correct result 
in the case of a measurement outcome. This implication is implicit 
in the extremisation procedure used to obtain the probability 
distribution (\ref{eq:maxent}). \par 

Given expression (\ref{eq:qsent}) for the quantum entropy, it is 
not readily obvious how $S_{\rho}$ depends on the eigenvalues of 
$\rho^{\alpha}_{\beta}$. In order to see this, all we require is 
the generating function $Z(\lambda)$ in (\ref{eq:gf}). As an 
illustration, let us consider the case of a two state system. We 
choose the basis where the density matrix is diagonal, with 
elements $\rho_{1}$ and $\rho_{2}=1-\rho_{1}$. Because 
$\lambda^{\alpha}_{\beta}$ commutes with 
$\rho^{\alpha}_{\beta}$, in this basis 
$\lambda^{\alpha}_{\beta}$ is also diagonal, with eigenvalues 
$\lambda_{1}$ and $\lambda_{2}$. The ${\sl\Gamma}$-space 
integration for the generating function $Z(\lambda)$ can be 
lifted to ${\bf C}^{2}$ with a spherical constraint on 
$\psi^{\alpha}(x)$. The integration involves a Gaussian (cf. 
\cite{dblh2}), and we obtain 
$Z(\lambda) = (2\pi)^{3}(e^{-\lambda_{2}} - e^{-\lambda_{1}})
(\lambda_{1}-\lambda_{2})^{-1}$, from which it follows that 
%%%%%%%%%%%%%%%%%%%%%%%%%%%%%18 
\begin{equation} 
\rho_{1}\ =\ \frac{1}{\lambda_{1}-\lambda_{2}} + 
\frac{1}{1 - e^{\lambda_{1}-\lambda_{2}}} . 
\end{equation} 
%%%%%%%%%%%%%%%%%%%%%%%%%%%%%% 
Then because the dependence on $\lambda^{\alpha}_{\beta}$ is only 
up to the eigenvalue difference, we can set $\lambda_{2}=\lambda$ 
and $\lambda_{1}=-\lambda$. With these expressions at hand, we can 
compare the quantum entropy with the von Neumann entropy. The 
qualitative behaviours of $S_{\rho}(\lambda)$ and 
$S_{\rm vN}(\lambda)$ in this example turn out to be similar, 
though not identical, as illustrated in Fig. 1, where we compare 
plots for the $\lambda$-derivatives of the two entropies. The two 
curves agree in the pure-state limits 
$\lambda\rightarrow\pm\infty$. \par 

%%%%%%%%%%%%%%%%%%%%%%%%%%%%% 
%%%%%%%%%%%%%%%%%%%%%%%%%%%%% 
\begin{figure}[b] 
%\label{} 
   \psfig{file=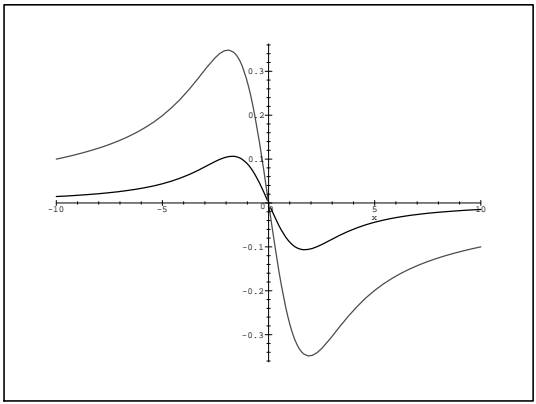,width=6cm,angle=270} 
 \caption{The plots for the entropy derivative 
$dS_{\rho}/d\lambda$ (lower-right curve) and the corresponding 
derivative $dS_{\rm vN}/d\lambda$ (upper-right curve) for the 
von Neumann entropy, for a two-state quantum system, showing 
that the two entropies have qualitatively similar behaviour.} 
\end{figure} 
%%%%%%%%%%%%%%%%%%%%%%%%%%%%% 
%%%%%%%%%%%%%%%%%%%%%%%%%%%%%  

Although we have only shown explicit results for a two-state 
system, it is worth remarking that the ${\sl\Gamma}$-space 
integration (\ref{eq:gf}) for the general generating function 
is invariably a Gaussian, and that the derivation of the entropy 
thus remains tractable for all finite dimensionalities. \par 

In summary, we have introduced the idea of a probability 
density function $\rho(x)$ on the space of rays through the 
origin of the Hilbert space that only reflects the information 
provided by the density matrix. Based upon this we were able to 
obtain the Shannon entropy for a quantum state, which, from 
an information theoretic point of view, is superior to von 
Neumann's proposal for the entropy. The utility of the 
distribution (\ref{eq:maxent}) does not exclusively reside, 
however, in studying the entropy of quantum states. In fact, 
it can be applied to numerous other probabilistic and 
information theoretic aspects of quantum mechanics, as well as 
quantum estimation theory. For example, the Lagrange multiplier 
$\lambda^{\alpha}_{\beta}$ in the foregoing analysis can be 
viewed as parameterising the quantum state 
$\rho^{\alpha}_{\beta}$ of the system. Then, in the problem 
of estimating an unknown quantum state \cite{helstrom}, it is 
of interest to consider the Fisher information matrix which 
determines the variance lower bound (cf. \cite{bc}). In the 
present context, 
this is given by the Hessian (\ref{eq:f-info}) of the generating 
function $\ln Z(\lambda)$, which can be computed explicitly for a 
given $\rho^{\alpha}_{\beta}$. The use of the minimal information 
state $\rho(x)$ can also be applied to the theory of quantum 
communication. We hope that the approach introduced here will 
offer further insights into the understanding of quantum 
theory. \par

DCB acknowledges PPARC for financial support. The authors 
acknowledge the Feza Gursey Institute of Istanbul for 
hospitality, where this work was carried out. Gratitude is 
expressed to B.K. Meister, Y. Nutku, and S. Popescu for 
stimulating discussion. \par

$*$ Electronic mail: d.brody@damtp.cam.ac.uk \par 
$\dagger$ Electronic mail: lane$\_$hughston@yahoo.com

\end{document}